\documentclass{article}
\usepackage[top=30truemm,bottom=30truemm,left=25truemm,right=25truemm]{geometry}
\usepackage{balance}
\usepackage{amsmath}
\usepackage{amsthm}
\usepackage{amssymb}
\usepackage{comment}
\usepackage{bm}
\usepackage[pdftex]{graphicx}

\title{A method to suppress local minima for symmetrical DOPO networks}
\author{Seiya Amoh\thanks{Tokushima University, 2--1 Minami-Josanjima,
    Tokushima, 770-8501 Japan} \and
Daisuke Ito\thanks{Gifu University, 1--1 Yanado, Gifu, 501-1193 Japan} \and
Tetsushi Ueta\footnotemark[1]}

\date{\empty}

\begin{document}
\maketitle

\begin{abstract}
  Coherent Ising machine (CIM) implemented by
  degenerate optical parametric oscillator (DOPO)
  networks can solve some combinatorial optimization problems.
  However, when the network structure has a certain type of symmetry,
  optimal solutions are not always
  detected since the search process may be trapped by local minima.
  In addition, a uniform pump rate for DOPOs in the conventional
  operation cannot overcome this problem.
  In this paper proposes a method to avoid trapping of
  the local minima by applying a control input in
  a pump rate of an appropriate node.
  This controller breaks the symmetrical property
  and causes to change the bifurcation structure
  temporarily, then it guides transient responses into the global minima.
  We show several numerical simulation results.
\end{abstract}

\section{Introduction}
In recent years, the combinatorial optimization problems appears
in various fields.

There are a lot of algorithms to tackle them, but many Non-deterministic
polynomial hard (NP-hard) problems require exponentially
increasing computation time as the problem becomes larger.

The Ising model is a mathematical formulation mimicking the ferromagnetism
dynamics, and the searching problem of its ground state
(namely the Ising problem\cite{Ising1925}) can encode
MAX-CUT problem\cite{Karp1972,Galluccio2000} for 3D network graphs.
The solution obtained from the coherent Ising machine (CIM)
takes a good value as
an approximate solution of the Ising problem.
It has been demonstrated that Ising problems
is solved by DOPO networks which is one of the implementations 
of CIM\cite{Wang2013}.
In the network, each DOPO corresponds to the Ising spin in
the problem, and the network state is obtained as a
solution when the pump rate exceeds the oscillation threshold.

Wang, et al. \cite{Wang2015} provided theoretical background
of the DOPO and showed the performance theoretical and
numerically, i.e., the coupled DOPOs can seek the optimal solution
as a ground state of the Ising model.
In most cases, the DOPO network gets the ground state
by gradually pumping all oscillators.
However, the solution of the DOPO network is not
always optimal\cite{Haribara2016}.
In the case of gradually pumping DOPOs, local minima,
which are not the ground state, appear earlier than the optimal solution,
therefore they disturb finding the ground state\cite{Takata2012,Takata2016}.
As a result, the performance of the CIM decreases.

Ito investigated bifurcation problems in a network configured by eight
DOPOs\cite{ITO201822}. By choosing a pump rate  of a specific node and
a unified pump rates of other nodes as a pair of parameters,
several bifurcation
diagrams composed by super-critical and sub-critical pitchfork
bifurcations and tangent bifurcations have been obtained.  The analysis of
their structures reveals that the variation of the unified pump rate
tends to be confined
into a valley sectioned by bifurcation curves. Inside of the valley,
the optimal solutions are concealed by other local minima.
This situation is caused by pitchfork bifurcations due to a symmetrical property,
thus we guess that the performance of the CIM may be improved
by breaking a symmetric property on purpose.
In this paper, we propose a method to suppress local minima by adjusting
not the unified pump rate but the specific one.
This control perturbs the symmetrical property of the DOPO network
temporarily and guides the trajectory into the
global solution. We show simulation results for several kinds of
symmetric DOPO networks.

\section{DOPO networks}
A DOPO system is described by the following $c$-number Langevin
equations\cite{Wang2013,Wang2015}:
\begin{equation}
  \begin{array}{rll}
    \displaystyle
    \frac{dc_j}{dt} &=& 
    \displaystyle
  	(p - 1 + (c^2_j - s^2_j))c_j + \sum_{l=1, l\not=j}^N \xi_{jl}c_l,\\
    \displaystyle
    \frac{ds_j}{dt} &=& 
    \displaystyle
    (p - 1 - (c^2_j - s^2_j))s_j + \sum_{l=1, l\not=j}^N \xi_{jl}s_l,\\
    && j = 1, 2, \ldots, N
  \end{array}
  \label{eqn:system}
\end{equation}
where $c$ and $s$ are the normalized in-phase and quadrature-phase
components, respectively. $p$ is the pump rate, $\xi$ is the coupling
coefficient of DOPO network.
A detailed description of the quantum models is shown in
Refs.\cite{Shoji2017b,Yamamura2017b}.
In the numerical simulations of Eq.~(\ref{eqn:system}),
we experimentally observe
that $s$ tend to be vanished as time goes on.
Thus the system can be simplified as follows:
\begin{equation}
  \frac{dc_j}{dt} = (p - 1)c_j - c^3_j + \sum_{l=1, l\not=j}^N \xi_{jl}c_l.
  \label{math:reduced}
\end{equation}
If we define the state vector $\bm{c}$ and edge weight matrix $\Xi$ as:
\begin{eqnarray}
  \bm{c} = \left(
  \begin{array}{c}
    c_1\\
    c_2\\
    \vdots\\
    c_N
  \end{array}
  \right),~~
  \Xi = \left(
  \begin{array}{cccc}
    0 & \xi_{12} & \ldots & \xi_{1N}\\
    \xi_{21} & 0 & & \xi_{2N}\\
    \vdots & & \ddots & \\
    \xi_{N1} & \xi_{2N} & & 0
  \end{array}
  \right),
\end{eqnarray}
where, $\xi_{jl}$ is a coupling coefficient between nodes j and l.
If node $i$ and $j$ is coupled, $\xi_{ij}=\xi_{ji}=\xi$, but
if not, $\xi_{ij}=\xi_{ji}=0$.
Then Eq. \ref{eqn:system} can be derived as Eq. (\ref{math:vectorized}).
\begin{eqnarray}
  \label{math:vectorized}
  \frac{d\bm{c}}{dt} = \bm{f}(\bm{c}) + \Xi\bm{c},
\end{eqnarray}
where
  $\bm{f}(\bm{c}) = (f(c_1), f(c_2), \ldots, f(c_N))^T$.
In Eq. (\ref{math:reduced}), the Jacobian matrix is derived as:
\begin{equation}
	J(\bm{c})=\frac{\partial \bm{f}}{\partial \bm{c}} +
	\frac{1}{\xi}\Xi.
\end{equation}

MAX-CUT problem is a problem that divides nodes of a graph into two
groups and maximizes the number of edges cut.
The Ising Hamiltonian can be applied to the number of
cuts in the MAX-CUT problem\cite{Galluccio2000}.
The number of cuts when describing the MAX-CUT problem in CIM
using DOPO networks is defined as:
\begin{eqnarray}
  \mathop{\rm cut}(\Xi, \chi) = 
  -\sum_{j=1}^N\sum_{l=1,l\not=j}^N J_{jl}\chi_j\chi_l,
\end{eqnarray}
where, $\chi_j = {c_j}/{|c_j|}$., $J_{jl}$ is the
generic element of the coupling matrix which defined as 
$J_{jl} = 0$ if $\xi_{jl} = 0$ , otherwise
$J_{jl} = 1$.
The relaxation function in Eq.~(\ref{math:rela}) is the smallest value
when the DOPO network is in the optimal solution.
\begin{equation}
	\eta(\bm{c})=\sum_{j=1}^{N}(p-c_j^2-1)=-\sum_{j=1}^{N}\sum_{l=1,
	l\ne j}^N\xi_{jl}\frac{c_l}{c_j}.
	\label{math:rela}
\end{equation}

\begin{figure}[htbp]
	\centering
	\begin{tabular}{cc}
    \includegraphics[width=0.3\hsize]{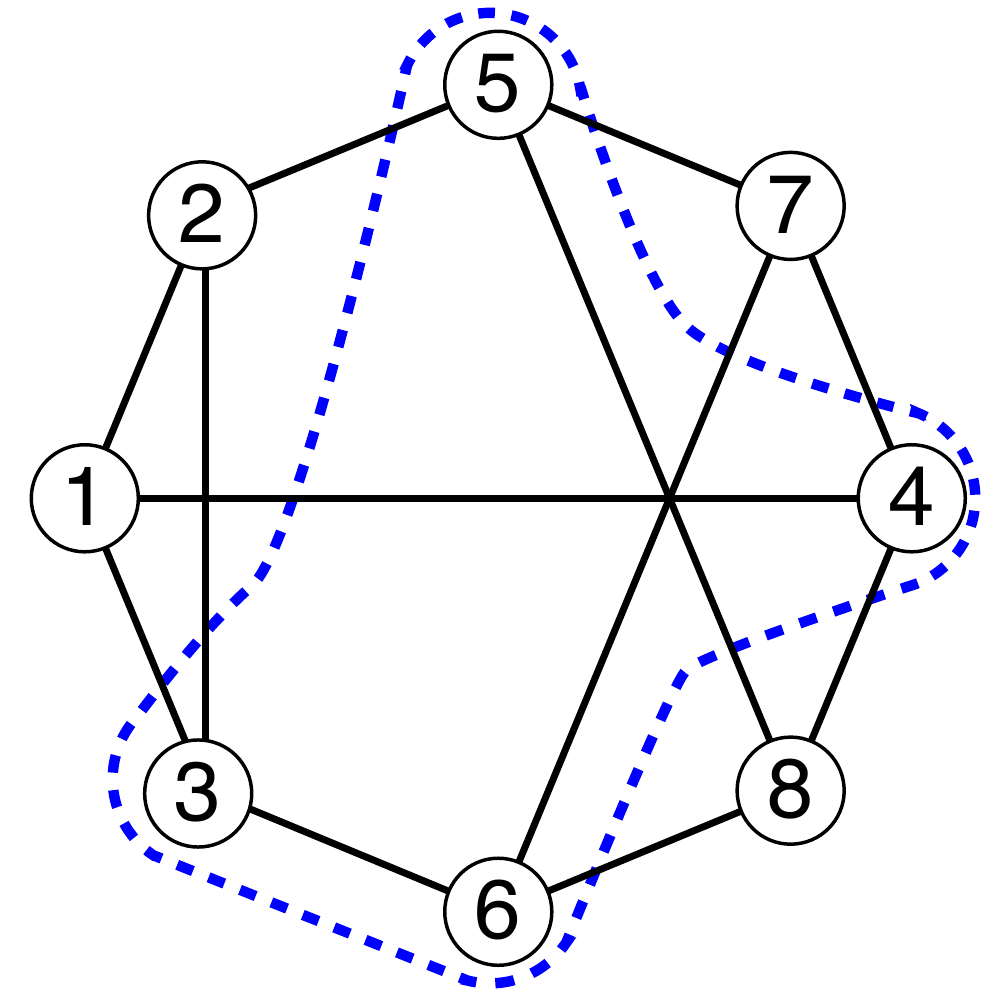} &
	  \includegraphics[width=0.3\hsize]{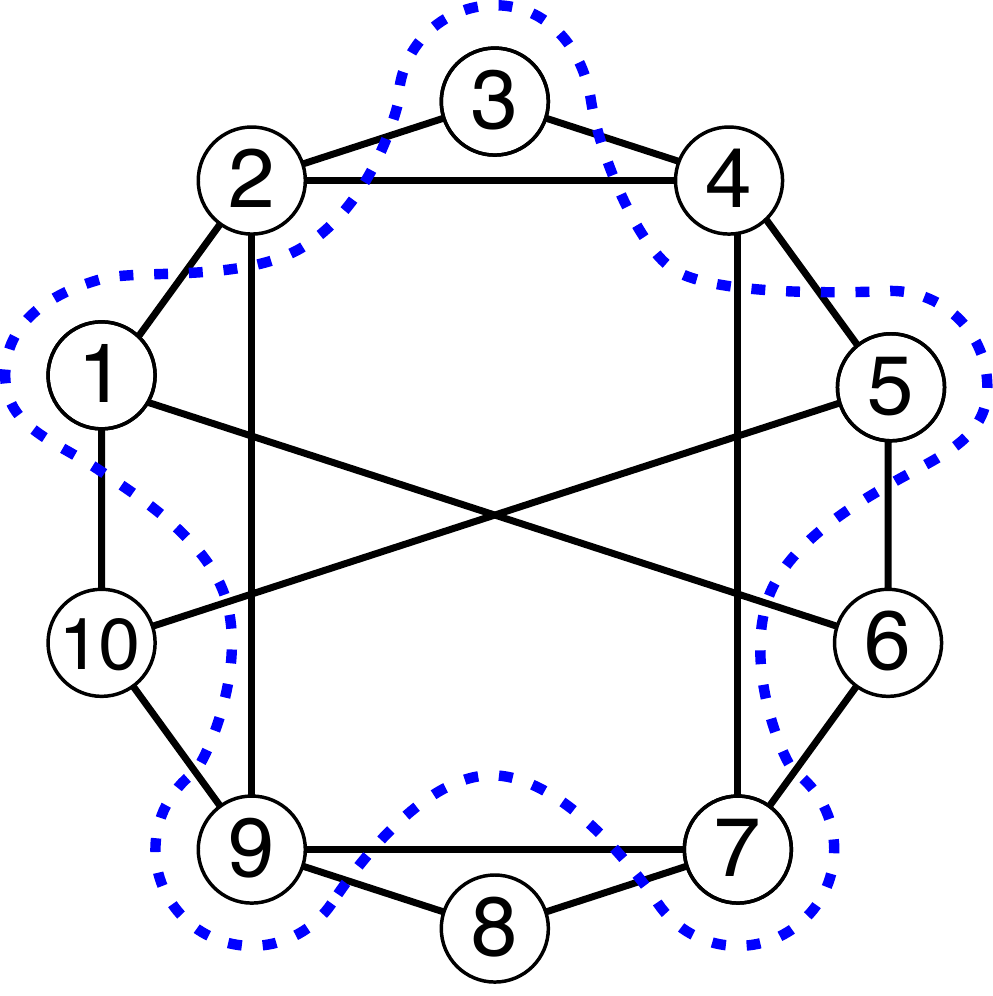}\\
	  (a) $N = 8$ & (b) $N = 10$
	\end{tabular}
	\caption{MAX-CUT problems and one of their optimal
	  solutions. The number of edges which the dashed line intersects give
	  the maximum for each graph.}
	\label{fig:n8n10}
\end{figure}

Figure \ref{fig:n8n10} gives MAX-CUT problems and their 
optimal solutions treated in this paper.
The cut is determined by the sign of the
solution $\bm{c}$ in Eq.~(\ref{math:reduced}), e.g.,
$c_1,c_2,c_7,c_8$ take positive value and
$c_3,c_4,c_5,c_6$ take negative value at Fig.~(\ref{fig:n8n10})(a).
Note that both graphs
contain certain classes of symmetry
and pitchfork bifurcations are the
major bifurcation phenomena of the DOPO networks\cite{Ueta2004}.

\section{Pump rate and its control}
\begin{figure}[h]
  \centering
  \begin{tabular}{cc}
    \includegraphics[width=0.31\hsize]{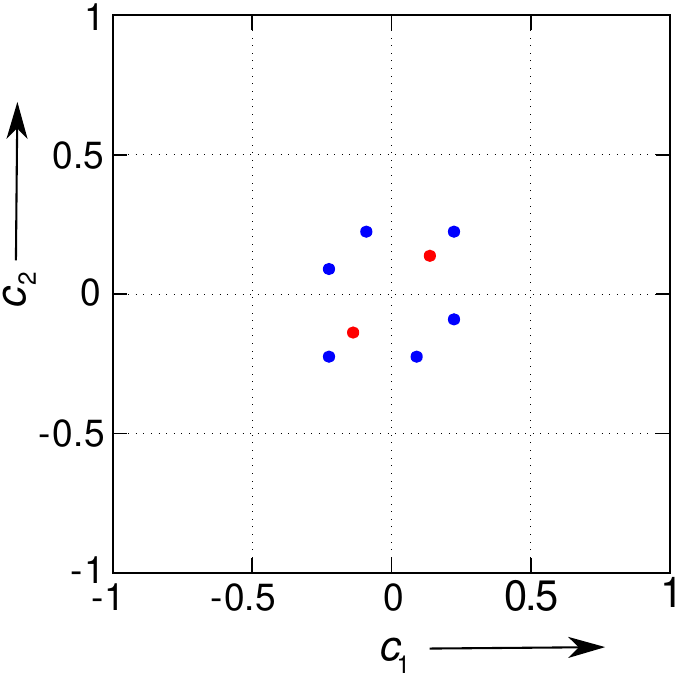} &
    \includegraphics[width=0.31\hsize]{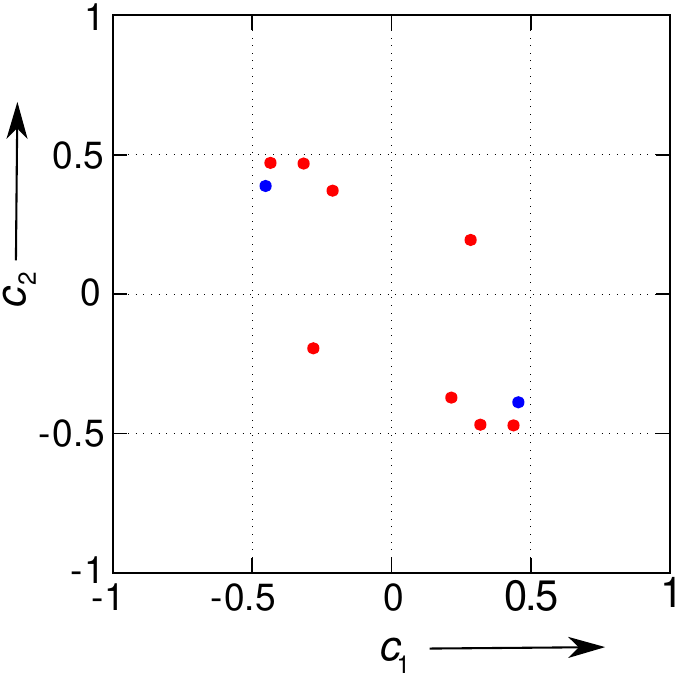}\\
    (a) & (b) 
  \end{tabular}
	\caption{ Stable equilibria. Blue points are the optimal solutions. 
	  (a): corresponding to Fig. \ref{fig:n8n10}(a),
	  (b): corresponding to Fig. \ref{fig:n8n10}(b).
	  Both are obtained with $p=0.92$.  
	}
  \label{fig:eqp8eqp10}
\end{figure}

We may obtain candidates of solutions of the problem as equilibrium points
of Eq.  (\ref{eqn:system}). From appropriate initial values,
the solution may fall
into one of stable equilibrium points according to both global solutions
and local minima after a transition state. 
To compute equilibrium points, we apply Newton's method by letting
the left-hand of Eq.~(\ref{math:reduced}) be zero,
with providing initial points by a brute-force strategy
with Mersenne twista\cite{MersenneTwista}.
Figure \ref{fig:eqp8eqp10}(a) and (b) are
stable equilibria of Eq. (\ref{math:reduced}) 
corresponding to Fig. \ref{fig:n8n10} (a) and (b), respectively.
There are many unstable equilibria from the pitch-fork bifurcation around
these attractors, but we do not visualize them.
Red and blue points are local minima and optimal solutions, respectively.
The pump rate is $p=0.92$ both of them.

The performance of the machine is improved by gradually increasing
$p$ from zero to one \cite{Takata2014} compared with the fixed pump rate.
That is, in the hardware implementation,
we practically increase the pump rate from
a low value, and gradually increase it near one monotonically.

\subsection{Bifurcations of equilibria with pump rates}
\label{sec:pump_rate_and_bifurcation}
\begin{figure}[htbp]
  \centering
    \includegraphics[width=0.58\hsize]{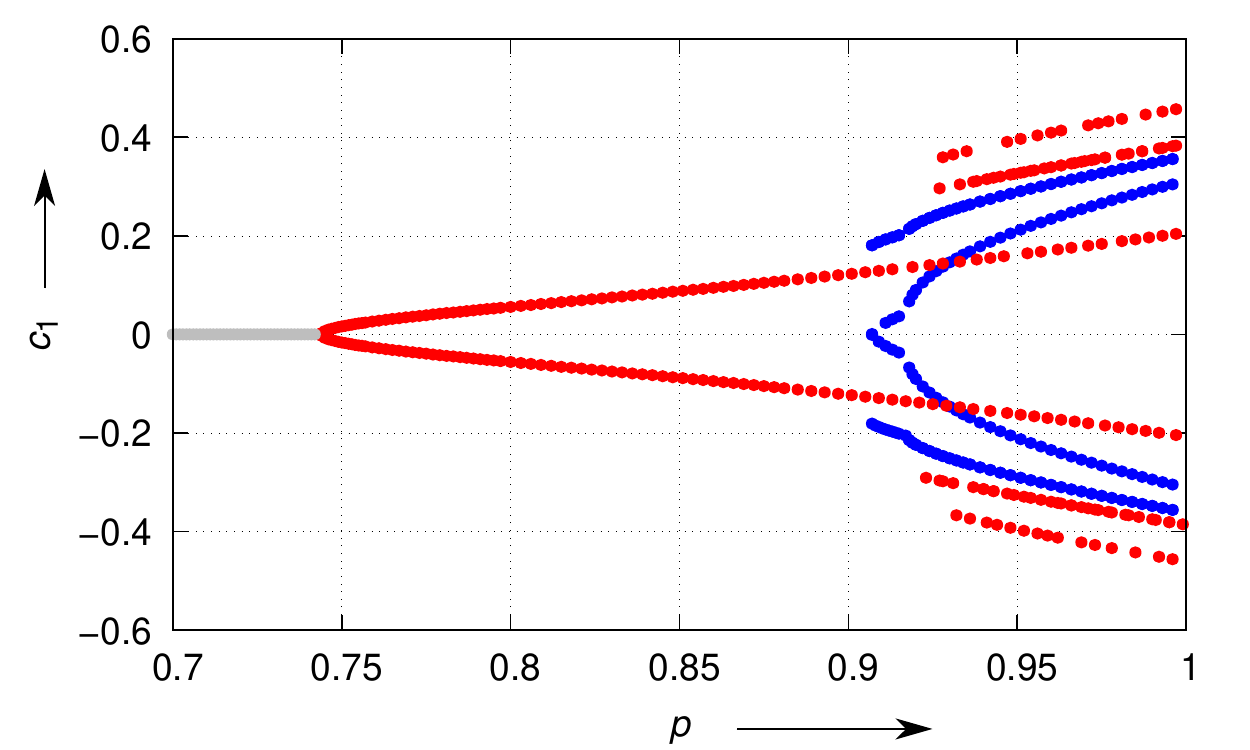} 
	\caption{One-parameter bifurcation diagram ($N=8$).}
  \label{fig:bif8}
\end{figure}

Figure \ref{fig:bif8} shows the one-parameter bifurcation diagram for
the case $N=8$ corresponding to Fig.~\ref{fig:n8n10}(a)
and this incremental pump rate process tracks this diagram
from left to right.
At any value of $p$, several random initial values
are scattered to find the equilibrium point.

Gray points show the origin stable point corresponding to no cut.
As $p$ increases, the system receives a pitch-fork bifurcation.
Therefore,
two symmetrical equilibrium points occur near the origin.
In this case, they are corresponding to local minima.
If $p$ increases further, and at $p = 0.909$, the system causes 
a multiple tangent bifurcation, and new four
equilibria appear. Indeed, they are optimal solutions (blue). 
Since local minima 
and optimal solutions coexist in $0.909 < p <  1$.
There is no parameter set
at which only the optimal solution occurs.
With this incremental pump rate process, local minima as sub-optimal
solution is detected at first, and the state is trapped to the same
solution against further increment of the pump rate.
It causes the performance deterioration of the machine.

\begin{figure}[h]
  \centering
  \includegraphics[width=0.58\hsize]{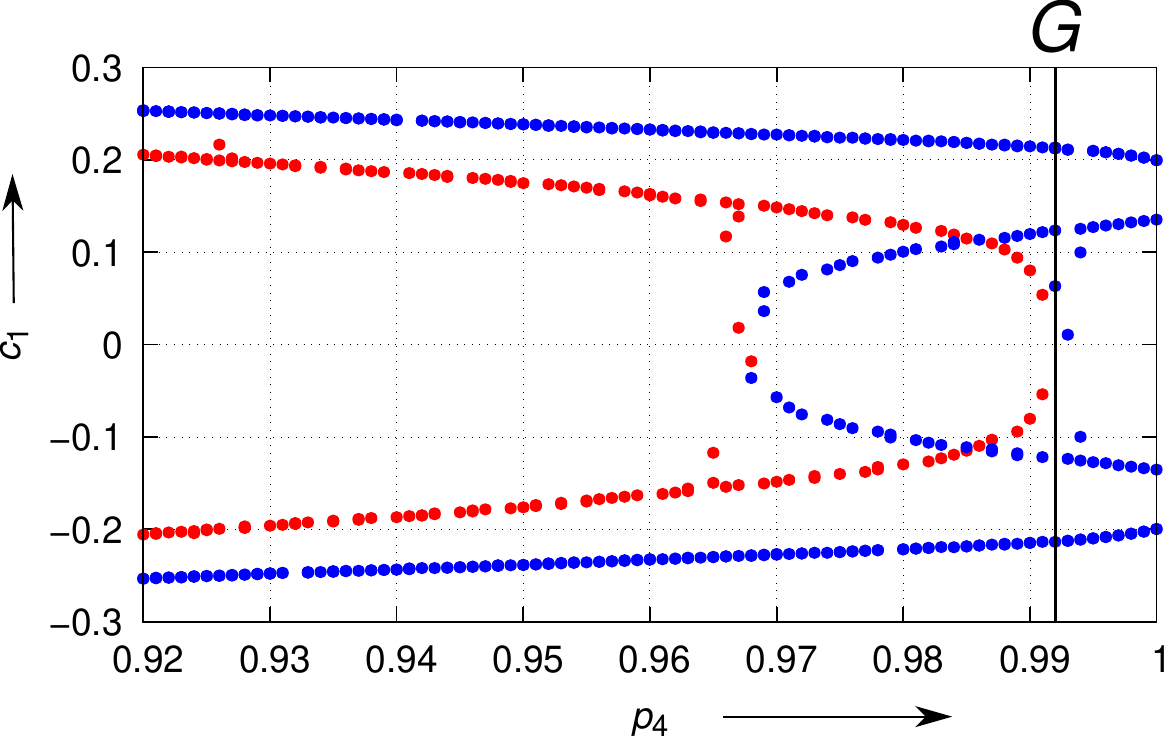}
  \caption{$c_1$-$p_4$ bifurcation diagram ($N=8$).}
  \label{fig:cuts8}
\end{figure}

Let $p_j$ be an individual pump rate
according to the node $j$ and assume it is adjustable.
Also assume that and other pump rates are fixed.
Figure \ref{fig:cuts8} shows an incremental process of $p_4$
for Fig. 1(a). The vertical axis shows the amplitude of
$c_1$. We increase $p_4$ from 0.9 to 1 gradually.
Blue and red point show optimal solutions and local minima, respectively.
The red solution meets a tangent bifurcation ($G$)
near 0.99, and then it disappears. 
Thus, there is parameter sets $0.99 < p_4$ that gives only optimal solutions by 
adjusting the pump rate individually.

\begin{figure}[h]
  \centering
  \includegraphics[width=0.4\hsize]{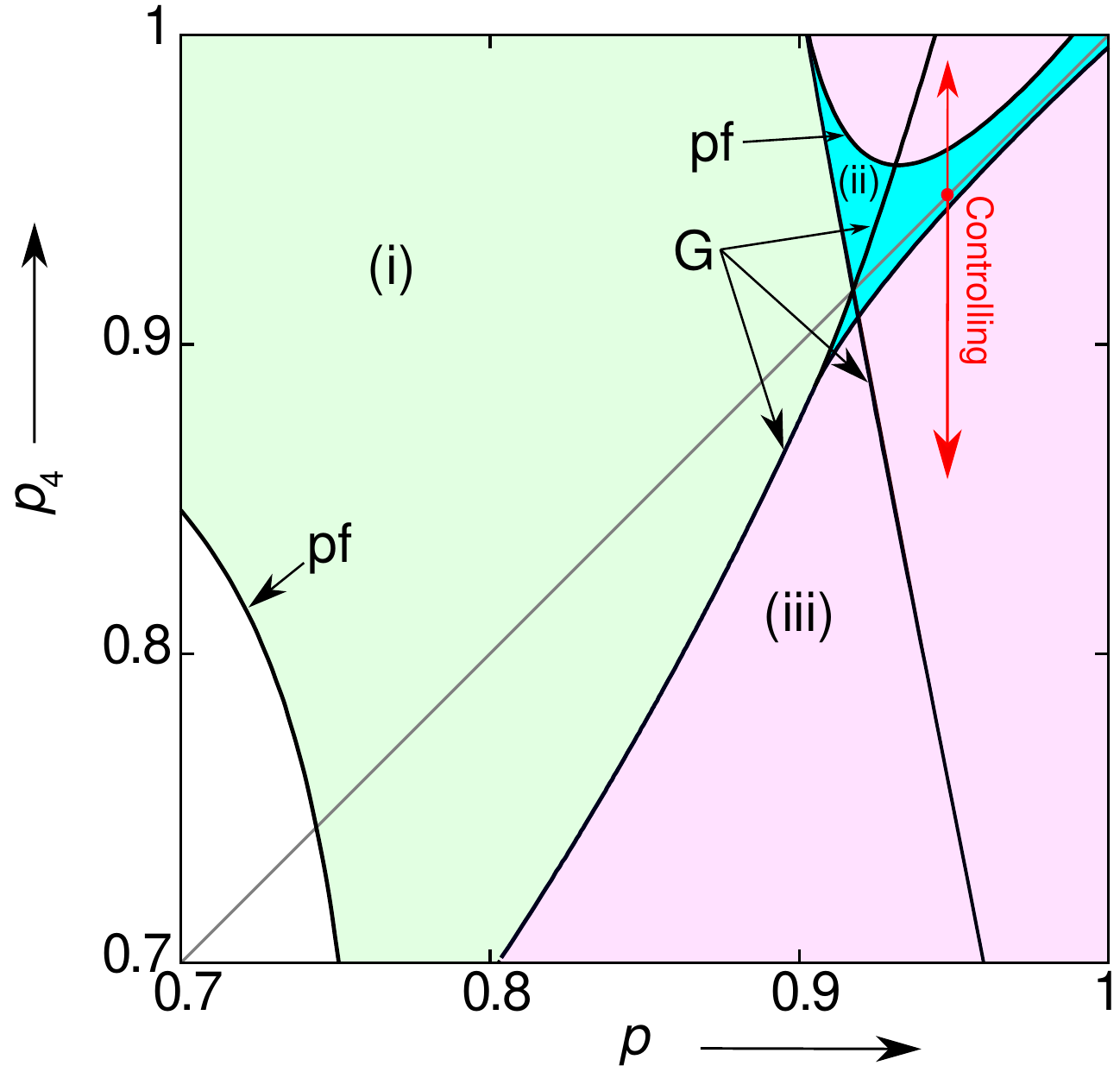}
  \caption{$p$-$p_j$ bifurcation diagram.}
  \label{fig:bif2dim}
\end{figure}

Now we investigate a relationship between optimal solutions and the pump
rate. As mentioned above, not only global minima but also
local minima are generated by pitchfork
bifurcations induced from the symmetrical configuration of the network.
If the system is trapped into local minima, 
the value of the relaxation Eq.~(\ref{math:rela}) becomes high
since this stable situation is given by the forcing energy frustration
to a part of nodes\cite{Haribara2016}. 
While, the precede work\cite{ITO201822} suggested that a
possibility of the control method to avoid trapping
to local minima by adjusting pump rates of some DOPOs.
This suppresses some bifurcation phenomena by breaking the symmetry of the
dynamical system on purpose. It can be called as a ``bifurcation
avoidance control'' so to speak.
Figure \ref{fig:bif2dim} shows the $p$-$p_j$ bifurcation diagram
at $N=8$,
where $G$ and $pf$ shows the tangent bifurcation
and pitchfork bifurcation, respectively.
We pick up a pump rate $p_j$ for one of DOPOs, and suppose a uniform
pump rate $p$ for other DOPOs.  
Only local minima occur in (i), both local minima and the optimal solutions
are mixed in (ii), and only optimal solutions occur in (iii).

On the gray diagonal line in Fig.~\ref{fig:bif2dim},
the pump rate set does not reach (iii).
However, controlling $p_4$ independently, as indicated by the red arrow,
the pump rate set reaches (iii).

\subsection{Pump rate control scheme}
In the Sect.~\ref{sec:pump_rate_and_bifurcation}, we recognize that
there is the pump rate set in
which only the optimal solution can be obtained by setting the pump rate
$p_j$ independently.
Although, with the method, it is necessary to specify the
amount to raise the pump rate in advance.
The pump rate set which is
only the optimal solution can be obtained from the 
relation with $p_j$ and $c_j$.
The frustration at a local minima imposed on nodes with small $|c_j|$.
To increase the $|c_j|$, increase the pump rate $p_j$ of a node with
the smallest $|c_j|$.
Therefore, consider a function that automatically adjusts $p_j$
according to the size of $|c_j|$,
and replace $p_j$ with the pump rate of the node
whose absolute value is small.
Assume that a variable range of the pump rate is from zero to one,
then $e^{-|c_j|}$ can be adopted as a reasonable adjustment control value.
As a result, the pump rate $p_j$ can be
regarded as a function depending on the state variable,
and the parameter can be set automatically.

\begin{figure}[h]
  \centering
  \includegraphics[width=0.5\hsize]{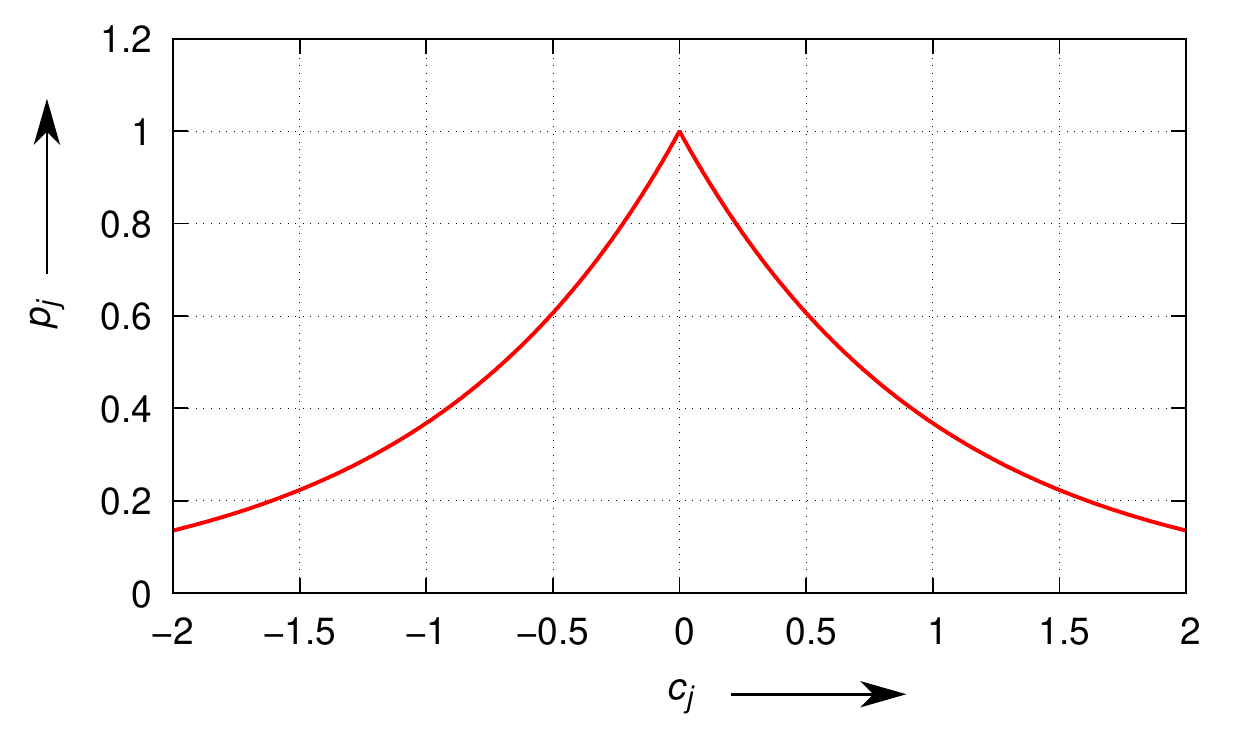}
  \caption{The example graph of Eq.~(\ref{eqn:control}), where $\beta = 1$.}
  \label{fig:exp}
\end{figure}

We propose a dynamic pump rate $p_j$ described as follows:
\begin{equation}
	p_j = \beta e^{-|c_j|},
  \label{eqn:control}
\end{equation}
where, $\beta$ is gain of control. In this paper we set $\beta = 1$.
When $c_j$ takes
a smaller value, $p_j$ inflates.
Figure~\ref{fig:exp} is the change of $p_j$.
In the case of Fig.~\ref{fig:bif2dim},
$p_4$ is controlled since $|c_4|$ is the smallest.
$p_4$ rises according to the exponential function of Fig.~\ref{fig:exp}.
As a result, the pump rate set moves to (iii).

\begin{figure}[h]
    \centering
    \includegraphics[width=0.23\vsize]{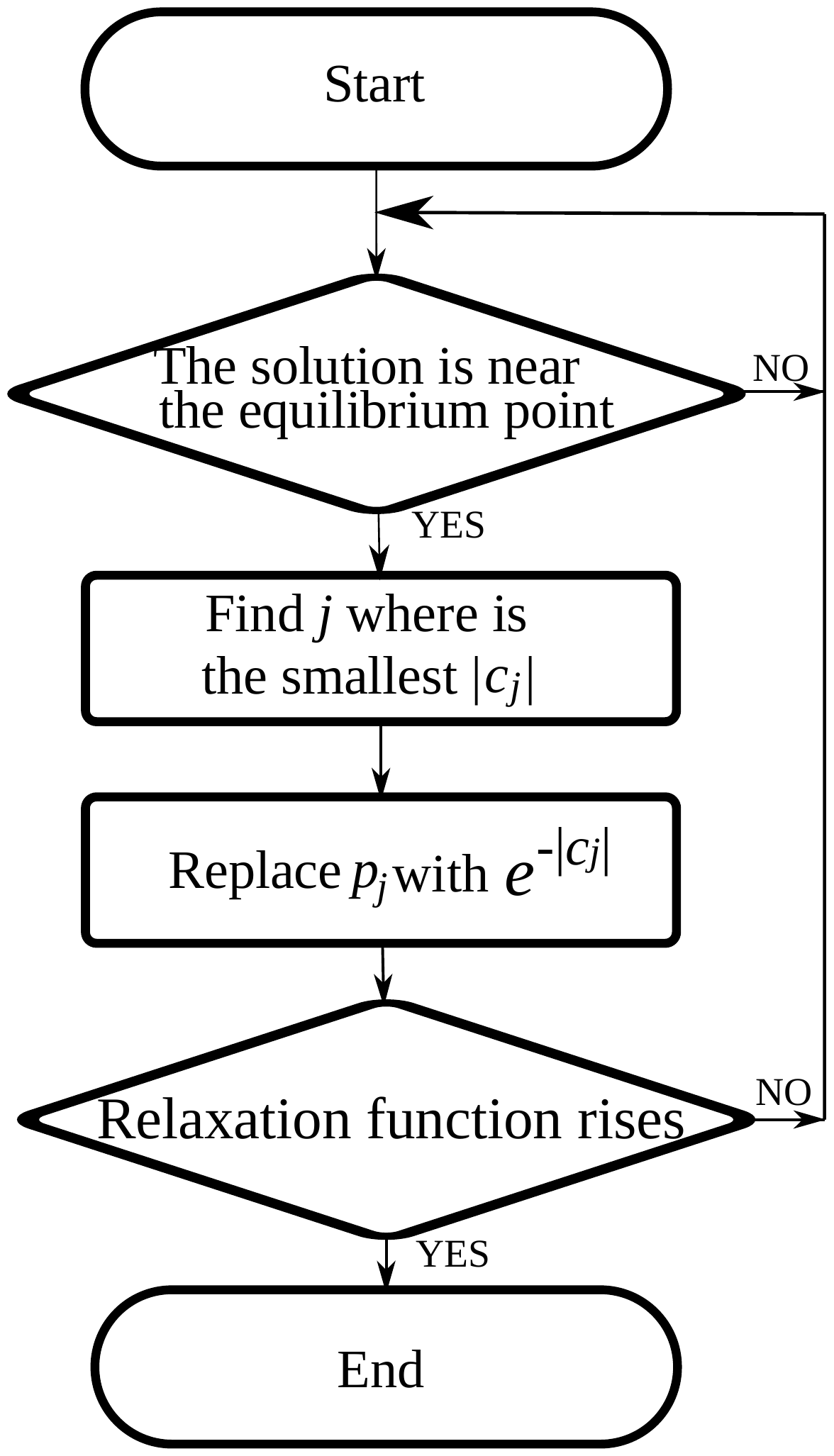}
    \caption{The flowchart of the control.}
    \label{fig:flowchart}
\end{figure}

Figure~\ref{fig:flowchart} depicts the procedure of the control.
When the calculation starts and the system becomes stable
near the equilibrium point, find the index $j$ of the
node with the smallest $|c_j|$.
Then, replace the pump rate of the $j$-th node with $e^{-|c_j|}$.
The number of replacing pump rate is not specified in particular,
excessive control causes the performance degradation,
so it is necessary to add control while monitoring the value
of the relaxation function.
In the following trials, we apply this procedure once.

\section{Simulation results}
\begin{figure}[h]
  \centering
  \includegraphics[width=0.43\hsize]{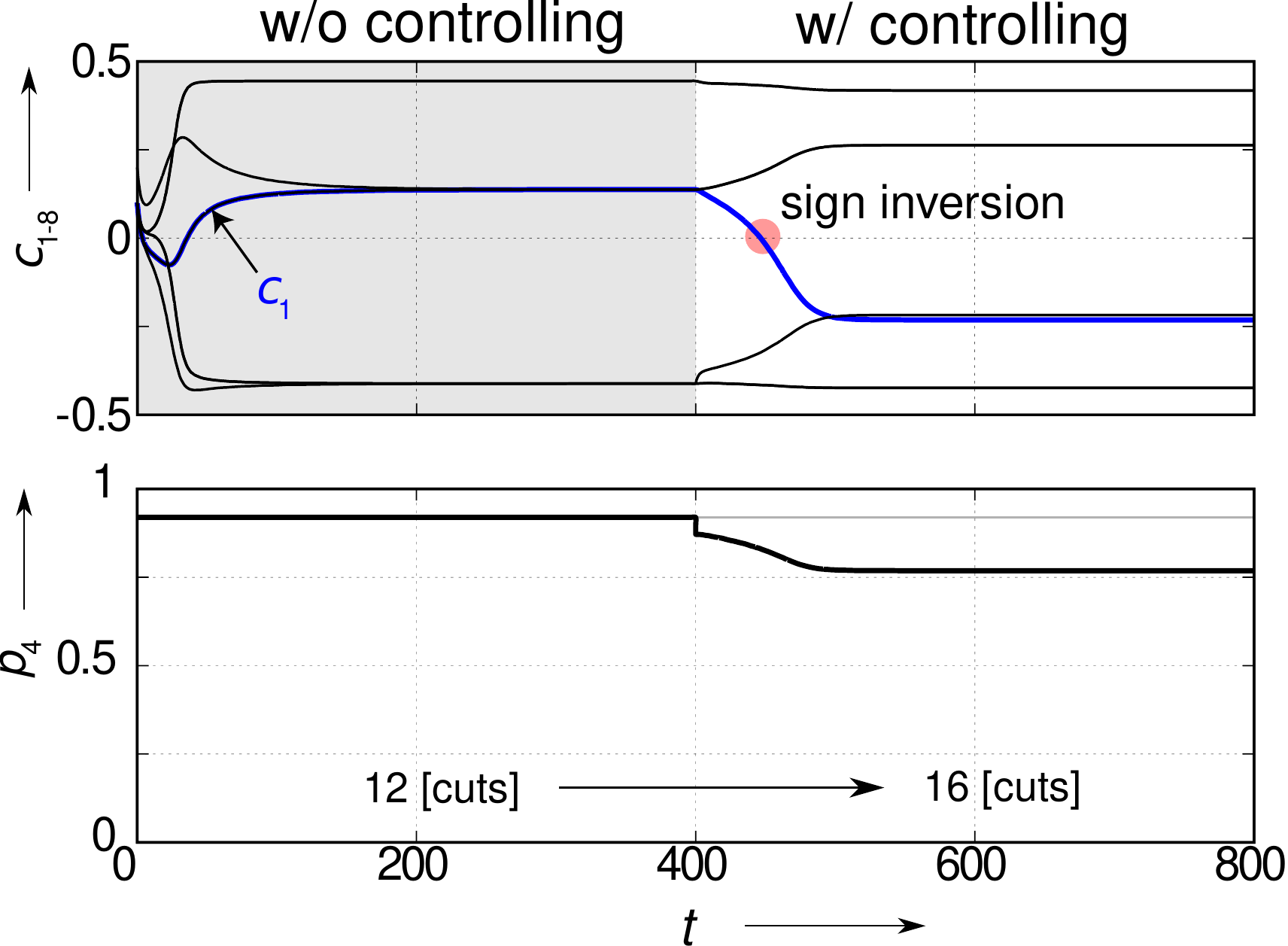}
  \caption{Control response for the $N=8$ network. The controller adjusts the
	  pump rate $p_4$.}
	\label{fig:uc8}
\end{figure}

Figure~\ref{fig:uc8} shows the controlling result for $N=8$ DOPO
networks shown in Fig.~\ref{fig:n8n10}(a) with the coupling
coefficient $\xi=-0.1$. We fix the unified pump rate as $p=0.92$.
As far as we checked, $|c_j|$ for all equilibrium points is
less than unity, thus we put $\beta = 1$.
The system runs without the controlling until
$t=400$, then it converges the local minima, and has
distribution of $c_{1-8}$ with large deviation.
In the case of the trial in Fig.~\ref{fig:uc8}, the node with $j=4$ is the
smallest $|c_j|$, so $p_4$ is selected as the control target.
The controlling starts after that, and the pump rate $p_4$ is derived from Eq.~(\ref{eqn:control}).
It is confirmed that the sign inversion of $c_1$ occurs
after $t=400$, and the number of cuts is improved from $12$ to $16$
by this sign inversion.
A red arrow in Fig.~\ref{fig:bif2dim} represents the change of $p_4$ in the parameter plane.
Note that, the local minima exist in the (i) region, the optimal and local minima coexist in the (ii) region.
In the (iii) region, the DOPO has only optimal solutions.
The parameter set moves from the blue region to the red region by the proposed controller
through the tangent bifurcation.
As a result, our controller can avoid the local minima.
By 100 trials with random initial values,
our controller improve the probability of finding the optimal solution
from $84\%$ to $100\%$.

\begin{figure}[h]
  \centering
  \includegraphics[width=0.43\hsize]{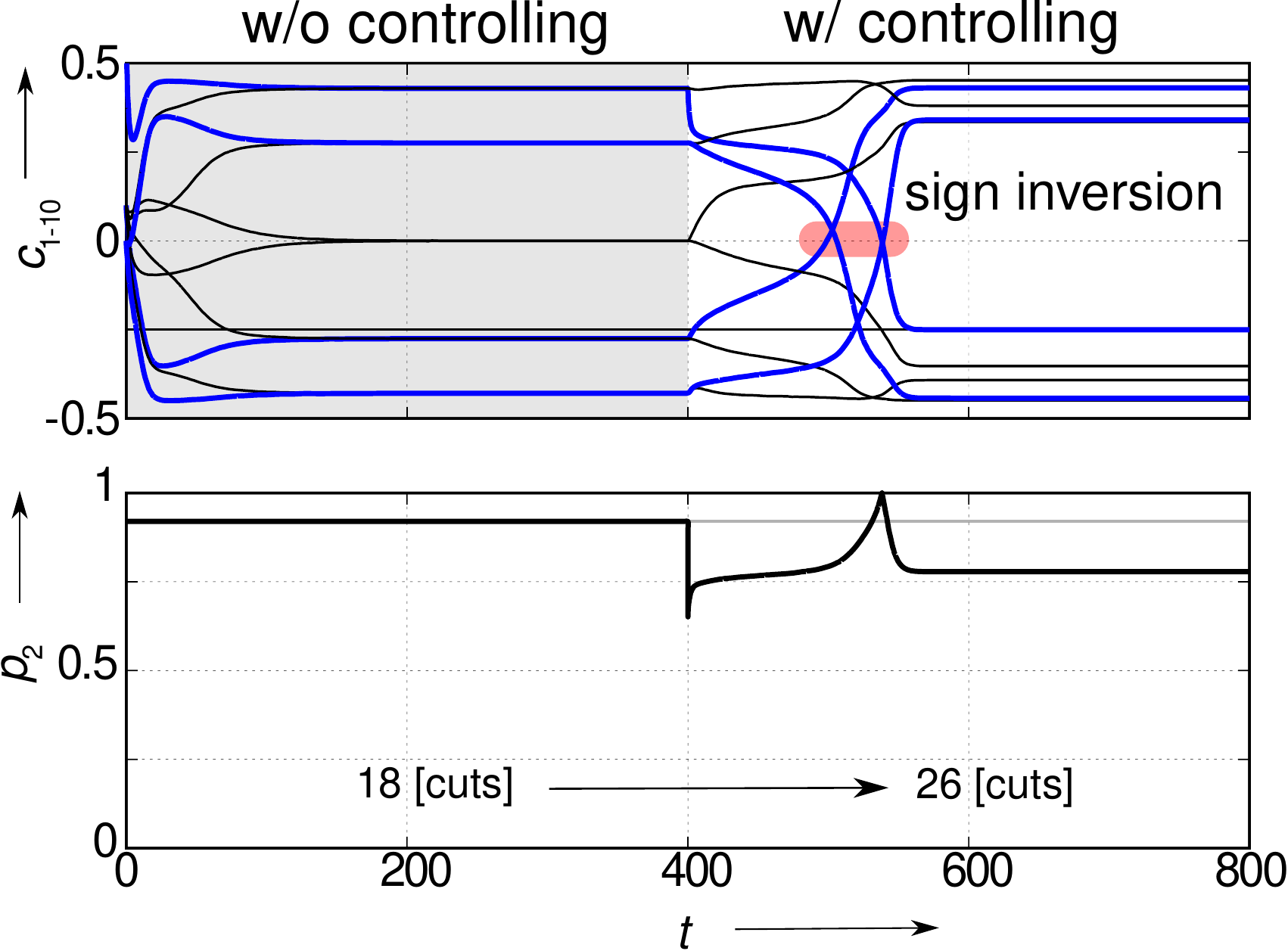}
  \caption{Control response for the $N=10$ network. The controller adjusts the
	  pump rate $p_2$.}
	\label{fig:uc10}
\end{figure}

The control response for the $N=10$ DOPO network is shown as
Fig.~\ref{fig:uc10}.
Here, the network structure is shown in Fig.~\ref{fig:n8n10}(b),
where the
coupling coefficient $\xi=-0.1$ and the common pump rate is $p=0.92$.
We choose the adjustable pump rate as $p_2$.
Similar to Fig.~\ref{fig:uc8}, the control improves the number of cuts
from $18$ to $26$.

\begin{figure}[h]
  \centering
  \begin{tabular}{cc}
    \includegraphics[width=0.22\hsize]{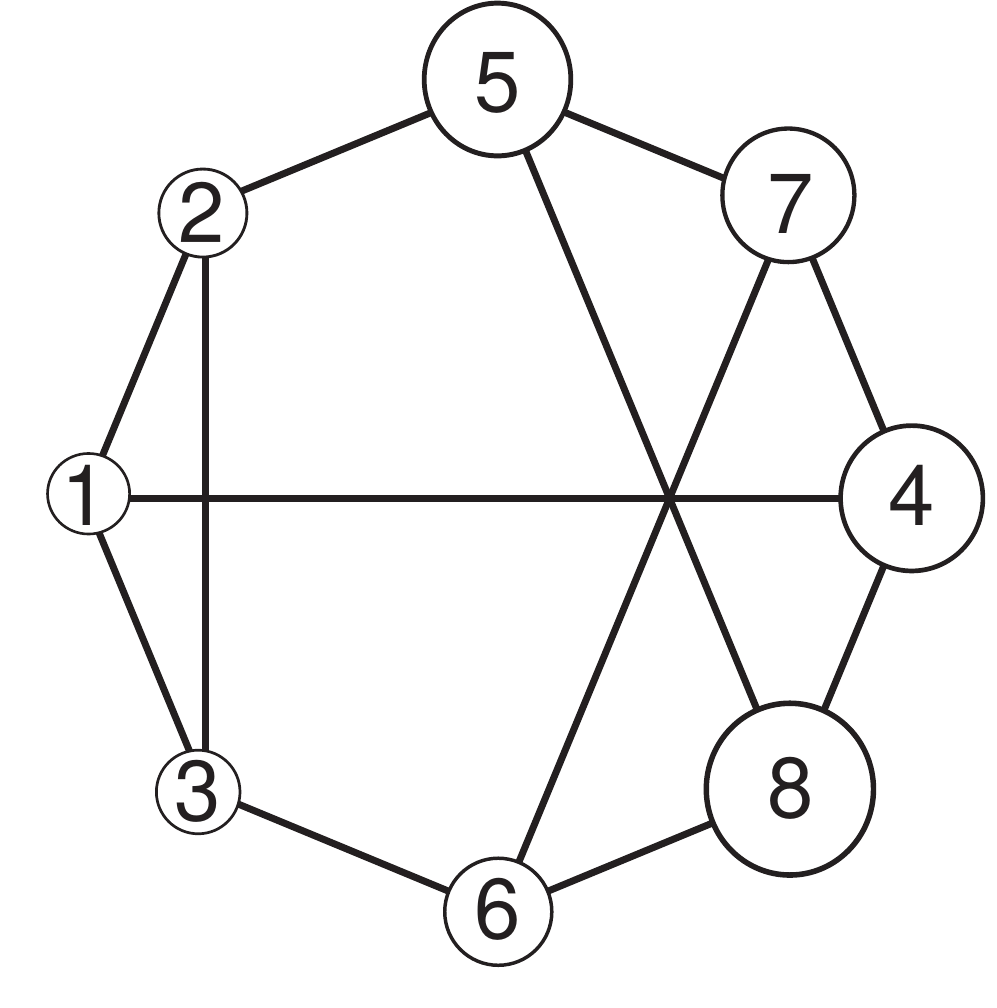} &
    \includegraphics[width=0.22\hsize]{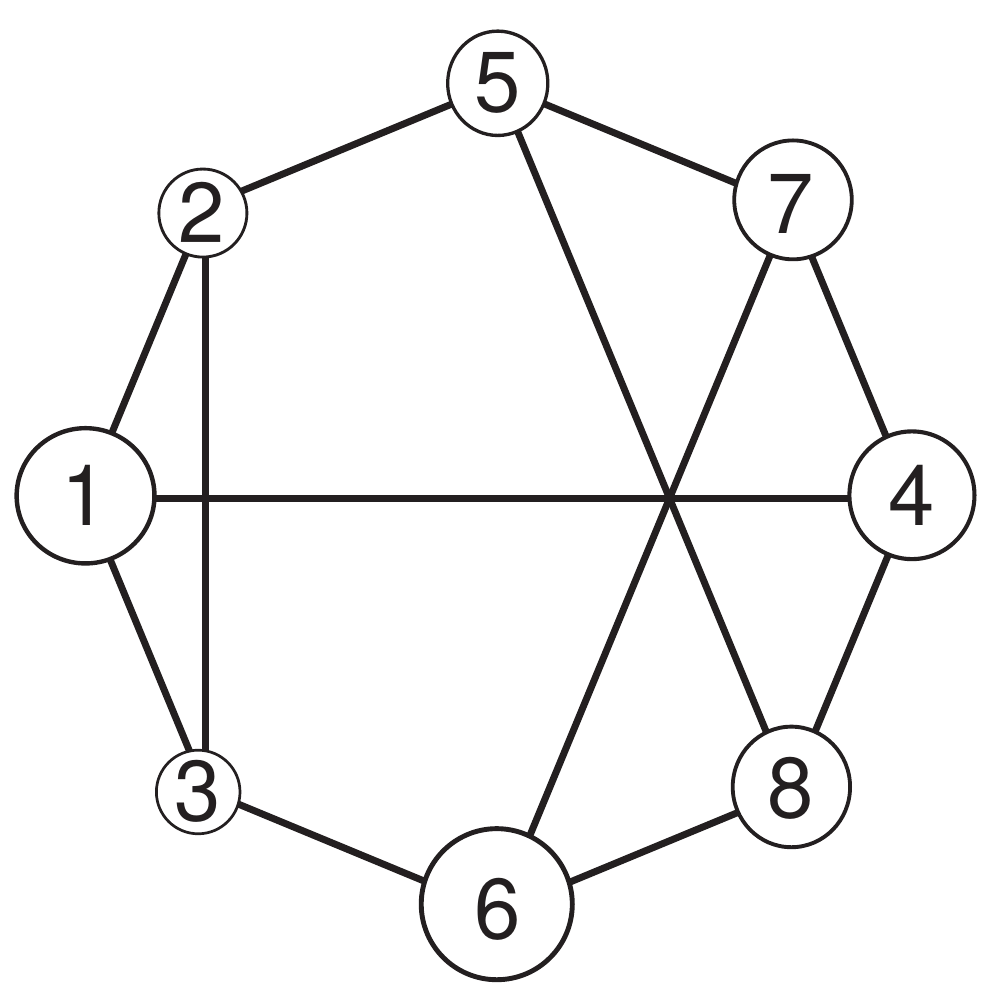}\\
    (a) & (b) 
  \end{tabular}
	\caption{
	  (a): before controlling,
	  (b): after controlling.
	}
  \label{fig:n8weight}
\end{figure}

Figure \ref{fig:n8weight} shows the relationship between eight nodes
on a steady-state with/without the proposed controller
corresponding to the simulation results in Fig.~\ref{fig:uc8}.
Note that, the size of each node is scaled by the absolute values of $c_j$
of the node $j$. 
When the network converges to local minima solutions without controlling,
the difference between maxima and minima circles becomes large;
therefore node five and eight indirectly frustrate node one.
Thus, they have $|c_j|$ values larger than other nodes.
However, the difference in the size of the nodes is
relieved because the network converges to the global optimal solution
from the local minima by the proposed controller.

\begin{figure}[h]
  \centering
  \begin{tabular}{cc}
      \includegraphics[width=0.25\hsize]{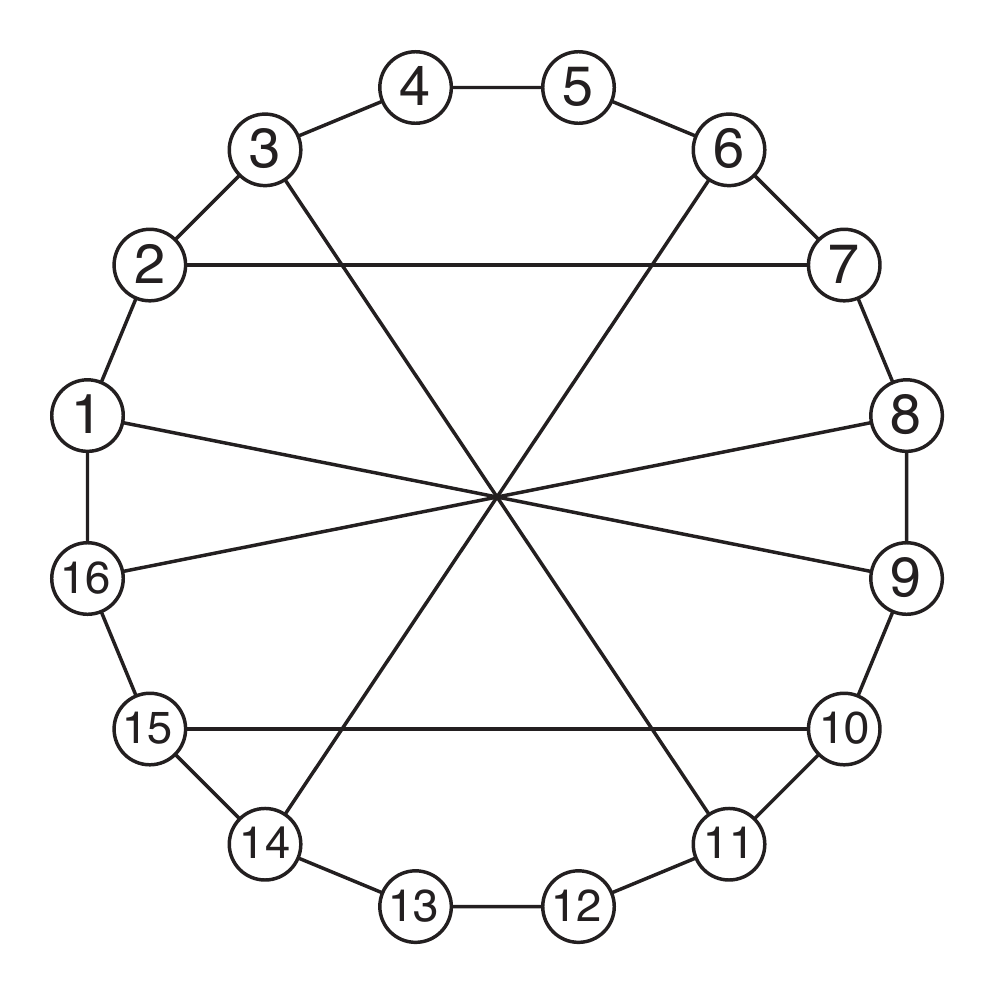}&
      \includegraphics[width=0.25\hsize]{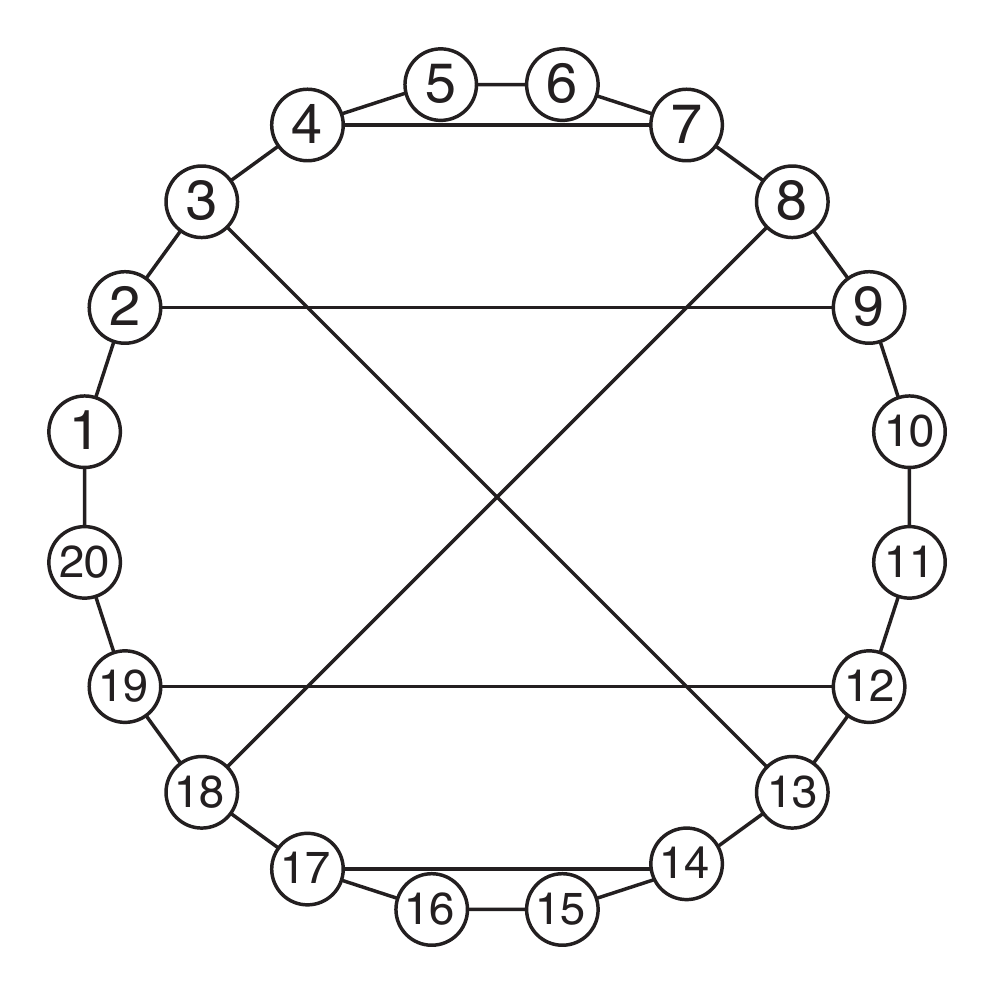}\\
      (a) $N = 16$ & (b) $N = 20$
  \end{tabular}
  \caption{The graph of $N=16$ and $N=20$.}
  \label{fig:n16n20}
\end{figure}

\begin{figure}[h]
  \centering
  \begin{tabular}{cc}
    \includegraphics[width=0.43\hsize]{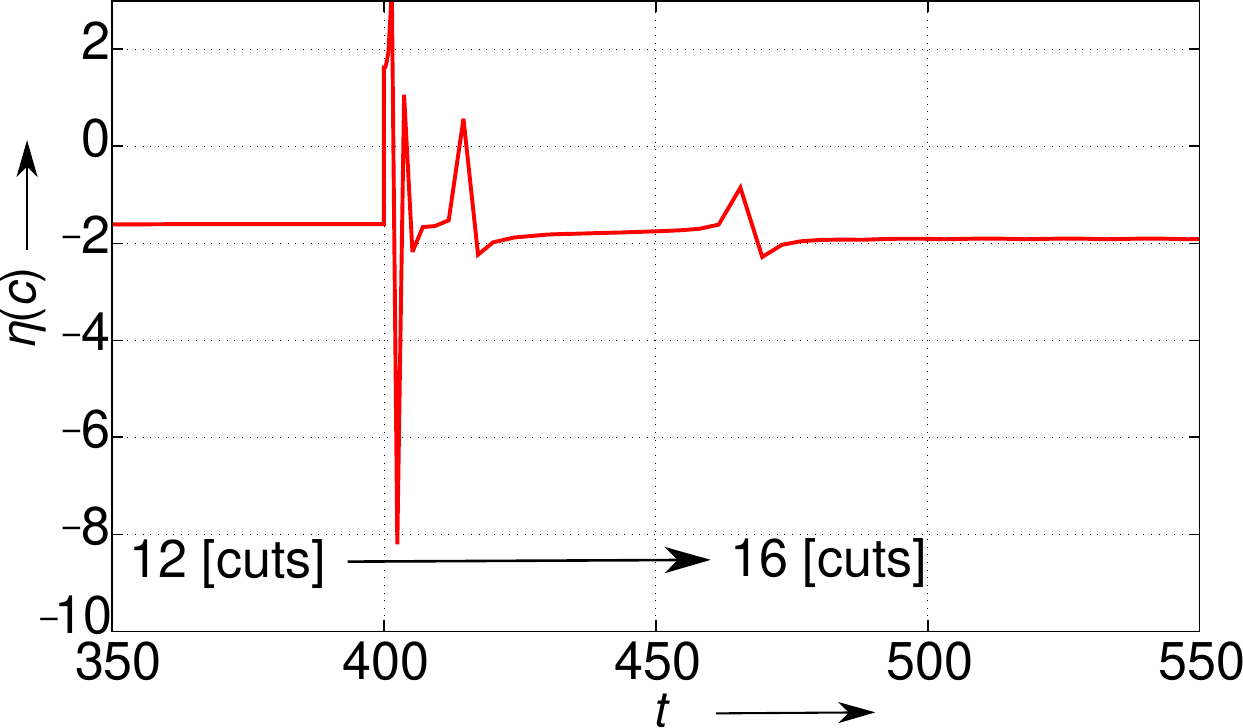} &
    \includegraphics[width=0.43\hsize]{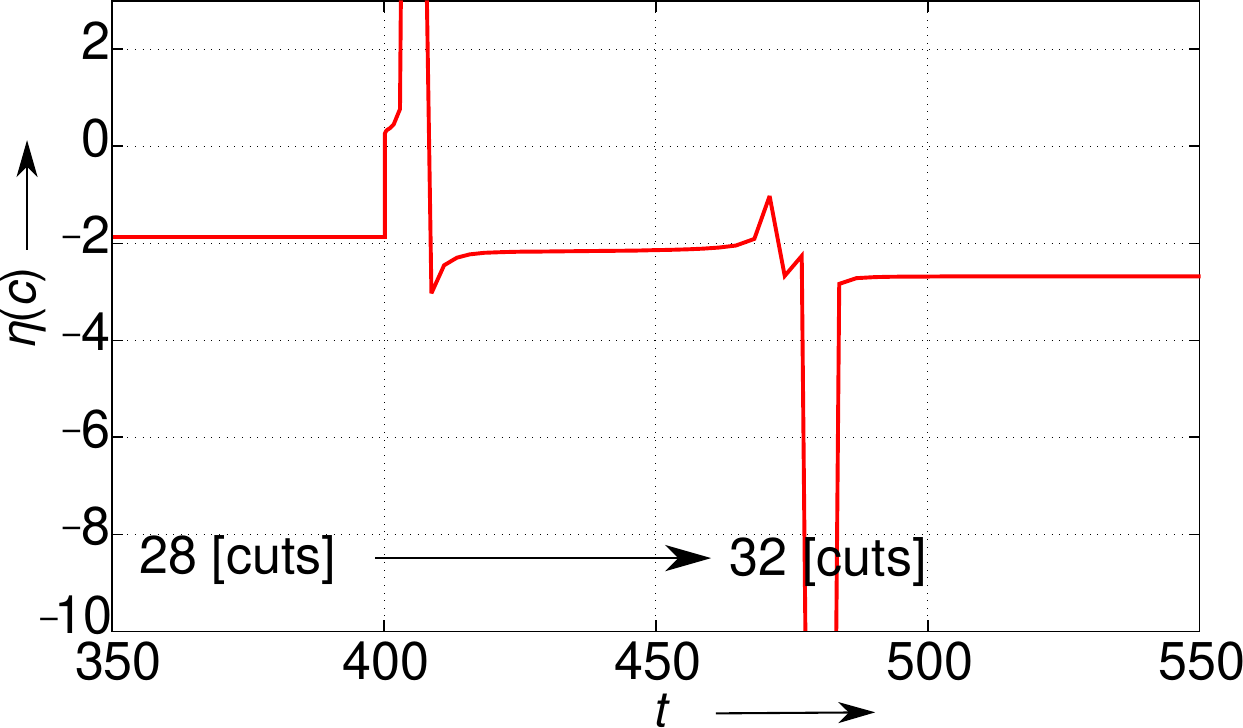}\\
    (a) & (b) \\
    \includegraphics[width=0.43\hsize]{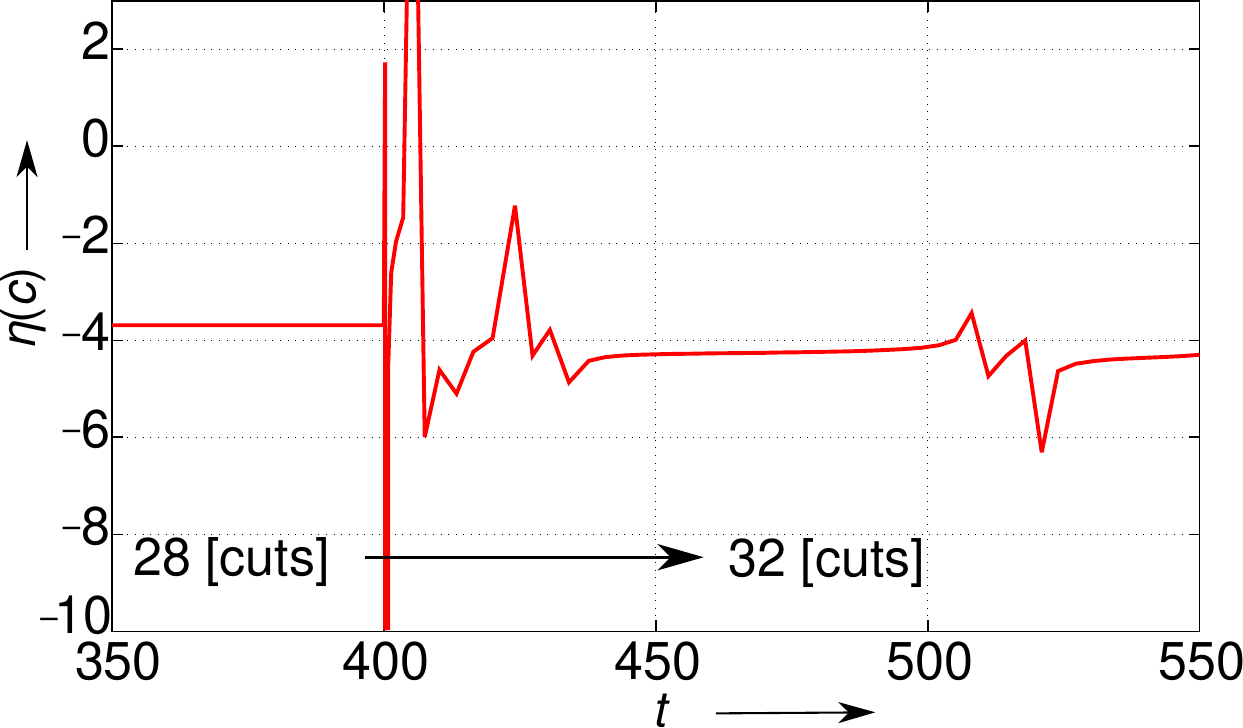} &
    \includegraphics[width=0.43\hsize]{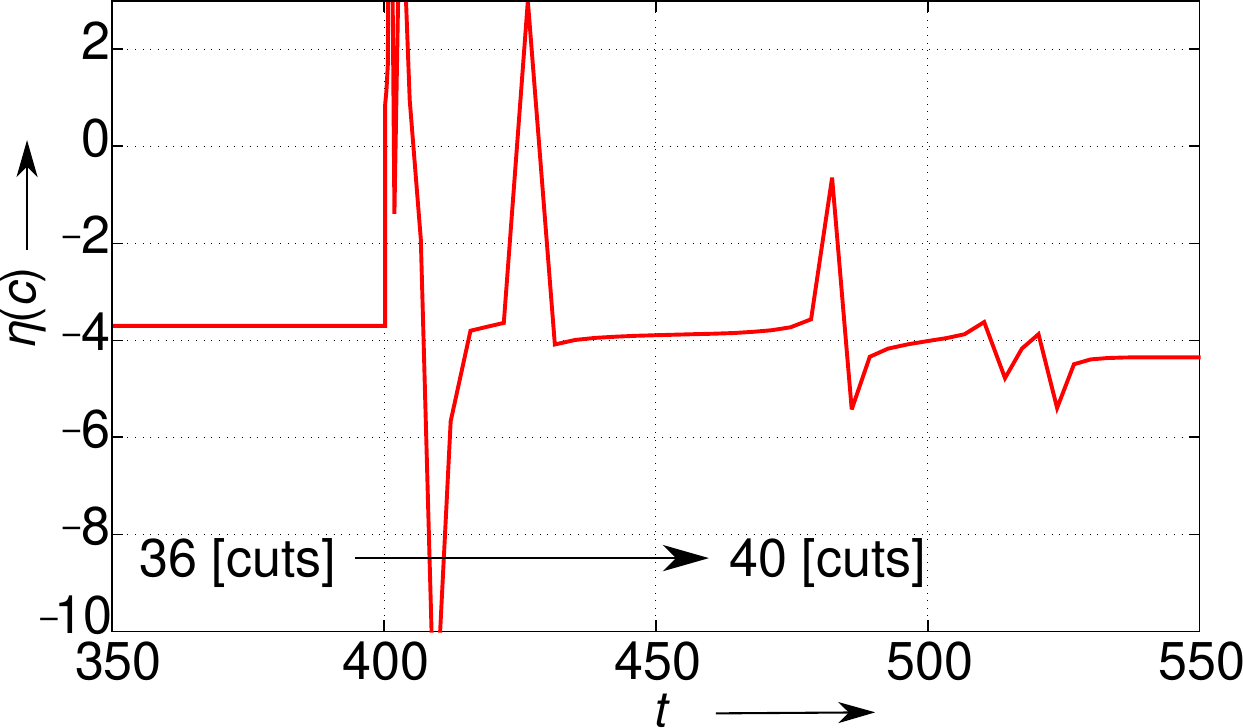}\\
    (c) & (d)
  \end{tabular}
	\caption{
    $\scriptstyle \mbox{Transitions of the relaxation function.}\atop
    \scriptstyle \mbox{
	  (a): $N=8$,
    (b): $N=10$,
    (c): $N=16$,
    (d): $N=20$.}$
	}
  \label{fig:relaxation_transition}
\end{figure}

Furthermore, Fig.~\ref{fig:relaxation_transition} shows the transition of the relaxation function in each graph.
The change of the node size in the graph of Fig.~\ref{fig:n8weight} is evaluated by Fig.~\ref{fig:relaxation_transition}(a).
The $\eta(\bm{c})$ of Fig.~\ref{fig:n8weight} decreases at $t=400$ due to the addition of control.
Fig.~\ref{fig:relaxation_transition}(c) and (d) are $N=16$ and $N=20$ nodes graph which are Fig.~\ref{fig:n16n20}.
Similar to Fig.~\ref{fig:n8n10}(a) and (b), Fig.~\ref{fig:n16n20} also have symmetry.
Initial pump rate of $N=16$ and $N=20$ are $p=0.98$.
It is found that the relaxation function decreased after control in all graphs,
and the number of cuts also increased.
The number of cuts for $N=16$ and $N=20$
improves from $28$ to $32$ and $36$ to $40$, respectively.

Table~\ref{tab:prob} summarizes the probability of
the optimal state with and without the proposed controller.
For the four networks, the proposed controller can
suppress the local minima, and can improve the probability of the optimal
solutions.

\begin{table}[h]
	\centering
	\caption{Probability of the optimal solution for w/o and w/ proposed controller.
	  The number of trials is 100. The initial state is given by random values.}
	\label{tab:prob}
	\begin{tabular}{c|c|c}
		number of nodes &w/o control& w/ control\\\hline\hline
		$8$& 84\%& 100\%\\
    $10$& 78\%& 100\%\\
    $16$& 92\%& 97\%\\
		$20$& 88\%& 100\%\\
	\end{tabular}
\end{table}

\section{Conclusion}
We propose a method to avoid local minima for
symmetrical DOPO networks.
We design new controlling scheme for the improvement of the
DOPO network as the MAX-CUT solver.
We summarize the proposal method and the results to following items.
\begin{enumerate}
\item{Conventionally, the pump rate was set uniformly,
  although we recognize that
  the local minima disappeared due to tangent bifurcation
  when the pump rate was set individually at each node.}
\item{There is a pump rate set that can obtain only the optimal solution
  in the $p_j$-$p$ bifurcation diagram of the DOPO network.}
\item{From the bifurcation diagram and the structure in the DOPO network,
  we proposed a controlling method that automatically sets the pump rate
  by the exponential function.}
\item{Our controller improves the probability of success for
  optimal solution search in the four symmetrical DOPO networks.}
\end{enumerate}
Next, the future tasks are follows:.
\begin{enumerate}
\item{In this paper, the relaxation function was monitored for the
  number of replacing pump rate,
  although we need to find out how to determine the
  most effective number of replacing.}
\item{We try to systematic explanations of the performance of
  proposed controller based on the bifurcation theory,
  however since there is a non-differentiable absolute value function
  in the control term, the bifurcation analysis becomes difficult.
  This problem can be avoided by replacing $e^{-|c_j|}$ with a
  quadratic function such as $1-{c_j}^2$, but the performance comparison
  between two control terms is required.}
\end{enumerate}

This paper is under article submission and will be published
on Nonlinear Theory and Its Applications, IEICE(Vol.E11-N, No.4, Oct. 2020).
All authors agree to post this preprint to arXiv.

\bibliographystyle{ieeetr}
\bibliography{paper}
\end{document}